\pdfoutput=1
\documentclass[11pt]{article}

\usepackage{graphicx}
\usepackage{amsmath}
\usepackage{dcolumn}
\usepackage{bm}
\usepackage{slashed}
\usepackage[bookmarks,bookmarksnumbered,colorlinks=true,anchorcolor=blue,
linkcolor=blue,urlcolor=blue,citecolor=blue,breaklinks=true]{hyperref}
\usepackage[utf8]{inputenc}
\usepackage{authblk}

\usepackage{color}
\usepackage{ulem}

\newcommand{\be}{\begin{equation}}
\newcommand{\ee}{\end{equation}}
\newcommand{\ba}{\begin{eqnarray}}
\newcommand{\ea}{\end{eqnarray}}
\def\bea{\begin{eqnarray}}
\def\eea{\end{eqnarray}}

\newcommand{\gsim}{\mathrel{\hbox{\rlap{\lower.55ex \hbox {$\sim$}}
                   \kern-.3em \raise.4ex \hbox{$>$}}}}
\newcommand{\lsim}{\mathrel{\hbox{\rlap{\lower.55ex \hbox {$\sim$}}
                   \kern-.3em \raise.4ex \hbox{$<$}}}}

\def\roughly#1{\mathrel{\raise.3ex\hbox{$#1$\kern-.75em%
\lower1ex\hbox{$\sim$}}}}
\def\lsim{\roughly<}
\def\gsim{\roughly>}

\def\({\left(}
\def\){\right)}
\def\[{\left[}
\def\]{\right]}
\def\<{\langle}
\def\>{\rangle}

\def\t{{\tau}}

\usepackage{xcolor}

\begin{document}
\title{\bf Holographic entanglement entropy of the double Wick rotated BTZ black hole}

\author[]{Mitsutoshi Fujita \thanks{fujita@mail.sysu.edu.cn}}
\author[]{Jun Zhang \thanks{zhangj626@mail2.sysu.edu.cn}}

\affil[]{\it School of Physics and Astronomy, Sun Yat-Sen University, Zhuhai 519082, China }


\maketitle

\begin{abstract}
In this paper, we analyze the holographic covariant entanglement entropy in the double Wick rotated version of a rotating BTZ black hole (3 dimensional Kerr-AdS solution), where the periodicity of Euclidean time and spatial direction are changed. {The dual field theory has negative energy in the Lorentzian signature.} The holographic entanglement entropy agrees with its CFT counterpart, which is obtained by a conformal transformation of the correlation functions of twisted operators. 
\end{abstract}

\newpage

\section{Introduction}
In contrast to correlation functions in quantum field theories, the entanglement entropy of a subsystem becomes a non-local quantity~\cite{Cardy,Review}. Dividing the space at a fixed time into two parts $A$ and $B$, the entanglement entropy is defined as
\ba
S_A=-\mbox{Tr}_A(\rho_A \log \rho_A),
\ea
where the reduced density matrix $\rho_A=\mbox{Tr}_B\rho$ is the trace of the total density matrix over the subregion $B$. For quantum critical phase transitions, the entanglement entropy diverges at the critical point and becomes an order parameter~\cite{Vidal:2002rm}. It captures geometric discernment of field theories such as an area law~\cite{EE}: the entanglement entropy defined in a subregion is dependent on properties of a shared boundary and resembles the black hole entropy. An area law also implies that the most entangled degrees of freedom come from the high energy ones located around an infinitesimal neighborhood of the entangling surface.

An interesting discovery was made by changing the periodicity of spacetime or adding boundaries~\cite{Affleck:1991tk,Cardy:1984bb}. When we express the entanglement entropy in the path integral formalism, the reduced density matrix and the entanglement entropy can be defined using the partition function on the $n$ copies of the manifold glued along $A$ at the fixed time. We consider Euclidean time and space direction.
When the subsystem $A$ is a single interval along a space direction with a periodic boundary condition, the entanglement entropy in 1+1 dimensional CFT becomes~\cite{Calabrese:2004eu}
\ba\label{PER1}
S_A=\dfrac{c}{3}\log \Big(\dfrac{R}{\pi a}\sin \Big(\dfrac{\pi l}{R}\Big)\Big)+a_1,
\ea
where $R$ is the circumference of the circle and $a_1$ is a constant, which is not universal. On the other hand, when the space direction is infinitely long,  entanglement entropy of a thermal mixed state at a finite temperature becomes 
\ba
S_A=\dfrac{c}{3}\log \Big(\dfrac{R}{\pi a}\sinh\Big(\dfrac{\pi l}{R}\Big)\Big)+a_1.
\ea
If two point functions of twisted operators are considered, this can be done via a conformal transformation. Originally, the cut was made into a circle. After other conformal transformations, the cut is along the axis of a cylinder. The entanglement entropy at a semi-infinite line or a finite $1d$ system with open boundaries has a constant term called boundary entropy, which does not depend on the length at zero temperature limit. The boundary entropy has the expression $S_0=\log g$, where $g$ is the ground state degeneracy. 

The Riemann surface can also have a general periodicity $z\sim z+2\pi \tau$, where $z$ is a Riemann surface complex coordinate. The density matrix then depends on the Hamiltonian and  momentum of the CFT as follows:
\ba
\rho =e^{-2\pi \tau_2 H+2 \pi i \tau_1 P}.
\ea
Entanglement entropy of CFT was computed with general periodicity~\cite{Hubeny:2007xt}, where the cut is along a spatial direction. The gravity dual of this CFT has angular momentum and becomes a rotating BTZ black hole, which is stationary but not static. The metric does not have a curvature singularity at a surface in the origin without matter couplings. This surface is a singularity in the causal structure. A rotating BTZ black hole has its Euclidean version with an analytic continuation of the angular momentum. 
 
Periodicity can be changed by using a coordinate transformation (the double Wick rotation) $z'=iz\equiv \sigma_1'+i\sigma_2'$ as follows:
\ba
z'\sim z' +2\pi (-\tau_2 +i\tau_1).
\ea
The motivation of this paper is to compute entanglement entropy with changed periodicity on both the CFT side and the gravity dual. The gravity dual is the double Wick rotated version of the rotating BTZ metric. It is interesting to see changes in the gravity dual because the rotating BTZ black hole is not static. It is necessary to analyze whether the metric is a black hole. Degrees of freedom in time-dependent backgrounds are also open problems. The entanglement entropy in $2d$ CFT is proportional to the central charge (degrees of freedom) and is useful.

{In this work, we compute the holographic entanglement entropy of the double Wick rotated spacetime and the CFT counterpart.  The Ryu-Takayanagi formula proposes the holographic dual of the entanglement entropy~\cite{Ryu:2006bv,Ryu:2006ef}, {which is the area of minimal surfaces (see also the review~\cite{Nishioka:2009un,Rangamani:2016dms}).} Entanglement entropy is a powerful tool to analyze strongly coupled systems. The hologrpahic entanglement entropy has been the order parameter of the confinement/deconfinement phase transition in a confining gauge theory~\cite{Nishioka:2006gr,Klebanov:2007ws,Buividovich:2008gq,Dudal:2016joz}.~\footnote{\cite{Jokela:2020wgs} shows that potential between quarks conveys more information than entanglement entropy~\cite{Jokela:2020wgs}.} A relation similar to the first law of thermodynamics is analyzed in \cite{Bhattacharya:2012mi,Guo:2013aca,Allahbakhshi:2013rda}.  
This relation was rewritten in terms of the relative entropy~\cite{Blanco:2013joa}. \cite{Hubeny:2007xt} proposed a covariant generalization of the holographic entanglement entropy, which is powerful to analyze the time-dependence in the dual field theory side. The area of the holographic covariant entanglement entropy (an extremal surface) can probe the interior of the black hole horizon, ending on a boundary~\cite{Hartman:2013qma}. The double Wick rotated version of the entanglement entropy is called geometric entropy. Holographic models are useful for computing geometric entropy. Using geometric entropy, \cite{Fujita:2008zv} analyzed the confinement/deconfinement phase transition of the Yang-Mills theory on compact space at finite temperature on both sides of duality.~\footnote{Third order phase transitions were captured by geometric entropy in free QCD-like theory with flavor on $S^1\times S^3$~\cite{Fujita:2010gx}. } Geometric entropy also played the role of the order parameter in $2d$ Yang-Mills theory~\cite{Fujita:2010gx,Gromov:2014kia}. The geometric entropy probed $AdS$ Schwarzschild black holes~\cite{Bah:2008cj} and the Reissner-Nordstrom $AdS$ background (effects of background charges)~\cite{Allahbakhshi:2013wk,Fujita:2020qvp}  by using a minimal surface. {However, the standard modular Hamiltonian is unusual in geometric entropy. A space direction is considered as the time in geometric entropy. By making use of the analytic continuation, one can define $\rho (t)$, which is the  density matrix of geometric entropy. One usually relies on the path integral formulation to compute geometric entropy.} Thus, the entanglement entropy is a good starting point to analyze spacetime with general periodicity.}

The rest of this paper is structured as follows. In Section 2, we review the rotating BTZ black hole. We derive the periodicity of Euclidean coordinates and the thermodynamics of the rotating BTZ black hole.  In Section 3, we analyze the double Wick rotated version of a rotating BTZ black hole. We obtain the periodicity of Euclidean coordinates via the double Wick rotation. In Section 4, we derive the holographic covariant entanglement entropy of the double Wick rotated spacetime. In Section 5, we derive the entanglement entropy in the dual CFT.

\section{The rotating BTZ black hole}

This section is the review of~\cite{Banados:1992wn,Carlip:1995qv}. We derive the periodicity of Euclidean coordinates in the rotating BTZ black hole by using coordinate transformations. We derive the stress energy tensor and free energy by using holographic renormalization. We analyze the thermodynamic properties of the rotating BTZ black hole, which describes theory at high temperature.  

The rotating BTZ metric becomes a solution of the $3d$ Einstein-Hilbert action as follows:
\ba
&ds^2=r^2 \Big(dx+\dfrac{r_+r_-}{r^2 L}dt\Big)^2+\dfrac{L^2 r^2 dr^2}{(r^2-r_+^2)(r^2-r_-^2)}-\dfrac{(r^2-r_-^2)(r^2-r_+^2)}{r^2 L^2}dt^2, \nonumber \\
& 8G_3 {M}=\dfrac{r_+^2+r_-^2}{L^2},\quad J=\dfrac{r_+r_-}{4 G_3L },
\ea
where $x\sim x+2\pi$ and $L$ denotes the AdS radius. Two integration constants are the mass $M$ and angular momentum $J$, which are conserved charges.  $J$ is related to rotational invariance. 

After the coordinate transformation {$(r,t)\to L (r,t)$} and $(r_+,r_-)\to L (r_+,r_-)$, coordinates become dimensionless. The metric is rewritten as
\ba
&ds^2=L^2\Big[r^2 \Big(dx+\dfrac{r_+r_-}{r^2 }dt\Big)^2+\dfrac{ r^2 dr^2}{(r^2-r_+^2)(r^2-r_-^2)}-\dfrac{(r^2-r_-^2)(r^2-r_+^2)}{r^2 }dt^2\Big], \label{MET2} \\
& 8G_3 M=r_+^2+r_-^2,\quad J=\dfrac{L r_+r_-}{4 G_3}, 
\ea 
Recall that the dimensions of $M,\ G_3^{-1},\ J/L$ are all 1.~\footnote{However, we keep $G_3$ because it is a dimensional quantity.} The blackening factor (or the lapse function) vanishes for two roots as follows:
\be
\begin{gathered}
  r_ + ^2 = 4{G_3}\left( {M + \sqrt {{M^2} - \frac{{{J^2}}}{{{L^2}}}} } \right), \hfill \\
  r_ - ^2 = 4{G_3}\left( {M - \sqrt {{M^2} - \frac{{{J^2}}}{{{L^2}}}} } \right). \hfill \\
\end{gathered}
\ee

{The Euclidean version $t\to -i\tau_E$ and $r_-\to - i\tilde{r}_-$ $(J\to - iJ_E)$ becomes}
\ba\label{MET8}
ds_E^2=L^2\Big[ r^2 \Big(dx-\dfrac{r_+\tilde{r}_-}{r^2}{d\tau_E}\Big)^2+\dfrac{r^2 dr^2}{(r^2-r_+^2)(r^2+\tilde{r}_-^2)}+\dfrac{(r^2+\tilde{r}_-^2)(r^2-r_+^2)}{r^2}d\tau_E ^2\Big].
\ea
The blackening factor vanishes when
{
\ba\label{RPM}
&r_+=2\Big[G_3M\Big(1+\Big(1+\dfrac{J_E^2}{M^2L^2}\Big)^\frac{1}{2}\Big)\Big]^{\frac{1}{2}},\nonumber \\
&\tilde{r}_-=2\Big[G_3 M\Big(-1+\Big(1+\dfrac{J_E^2}{M^2L^2}\Big)^\frac{1}{2}\Big)\Big]^{\frac{1}{2}}.
\ea}
These two values don't agree when $M$ and $J_E$ are finite. This difference shows that there are no extreme limits, unlike Lorentzian black holes.~\footnote{In a Lorentzian rotating black hole, {$J_E=i J$ and $\tilde{r}_-=ir_-$.} The extreme limit means $r_+=r_-$ when $J=ML$. Moreover, the black hole mass should be larger than the angular momentum $ML\ge J$ when the black hole horizon exists. }

The rotating BTZ metric in the Euclidean signature has constant negative curvature and the spacetime becomes locally equivalent to hyperbolic three-space. The following Euclidean coordinate transformation maps to a Poincare AdS metric
\ba
&X=r_1\cos\theta=\sqrt{\dfrac{r^2-r_+^2}{r^2+\tilde{r}_-^2}}\cos\Big(r_+\tau_E+\tilde{r}_-x\Big)\exp \Big(r_+ x -\tilde{r}_-\tau_E \Big), \nonumber \\
&Y=r_1\sin\theta=\sqrt{\dfrac{r^2-r_+^2}{r^2+\tilde{r}_-^2}}\sin\Big({r_+}\tau_E+\tilde{r}_- x\Big)\exp \Big({r_+} x -{\tilde{r}_-}\tau_E\Big), \nonumber \\
&z=\sqrt{\dfrac{r_+^2+\tilde{r}_-^2}{r^2+\tilde{r}_-^2}}\exp \Big({r_+} x -{\tilde{r}_-} \tau_E\Big),
\ea
where a spherical coordinate is represented by $\theta =r_+\tau_E  +\tilde{r}_-x$ and $r_1= \sqrt{\frac{r^2-r_+^2}{r^2+\tilde{r}_-^2}}\exp \Big(r_+x -\tilde{r}_-\tau_E \Big)$. When $r\to \infty$, coordinates mentioned above describe $(X,Y)$ plane at $z=0$. When $r\to r_+$, coordinates mentioned above describe the $z$ axis at $r_1=0$. The Euclidean metric is rewritten as
\ba
ds_E^2=\dfrac{L^2}{z^2}(dX^2+dY^2+dz^2)=\dfrac{L^2}{z^2}(dr_1^2+r_1^2d\theta^2+dz^2).
\ea

The periodicity of $x$ means
\ba
(r_1,\theta)\sim (r_1 e^{2\pi r_+},\theta +{2\pi \tilde{r}_-}),
\ea
where a translation is along a radial direction and a $2\pi \tilde{r}_-$ rotation is a rotation around the $z$ axis.

The periodicity of the trigonometric functions $(\theta\sim \theta+2\pi)$ and nonsingular transformation at the $z$ axis ($r=r_+$ and $r_1=0$) require
\ba\label{PER13}
&(x,\tau_E)\sim (x +\phi_0, \tau_E+\beta_0), \\
&\beta_0=\dfrac{1}{TL} =\dfrac{2\pi r_+}{r_+^2+\tilde{r}_-^2},\quad \phi_0=\dfrac{2\pi \tilde{r}_-}{r_+^2+\tilde{r}_-^2}.
\ea
If we don't require the periodicity mentioned above, the Euclidean black hole receives a conical singularity at the black hole horizon. 

We define the chemical potential as the shift of the rotating BTZ black hole $\Omega_E =\frac{4G_3J_E}{r_+^2L^2}=\frac{\tilde{r}_-}{r_+L}$. We then find the relationship 
\ba
\phi_0=\beta_0 \Omega_E L.
\ea 
We use a complex coordinate $z=z+i 2\pi \tau_E$. Periodicity of $z$ is given by
\ba
z\sim z+\beta_0 \Omega_E L+i\beta ,\quad \bar{z}\sim \bar{z}+\beta_0 \Omega_E L-i\beta.
\ea
Following the analytic continuation $\Omega =-i \Omega_E$, left and right temperature are as follows:
\ba\label{TEMLR}
z\sim z+i \beta_0 (1+L\Omega), \bar{z}\sim \bar{z} -i \beta_0 (1-L\Omega ).
\ea

We then compute the holographic stress energy tensor. 
To compute the holographic stress energy tensor, we perform the coordinate transformation and use the FG coordinate with $g_{\mu \rho}=0$ as follows: 
\ba ds^2=\frac{L^2}{\rho^2}(d\rho^2+\bar{g}_{ij}dx^idx^j), \ea
where 
\ba
\bar{g}_{ij}=\bar{g}_{ij}^{(0)}+\bar{g}_{ij}^{(2)}\rho^2+\bar{g}_{ij}^{(4)}\rho^4+\dots
\ea
The radial coordinate $\rho$ can be perturbatively represented by the large $r$ expansion. Metric for \eqref{MET2} becomes
\ba\label{MET13}
\bar{g}_{tt}^{(0)}=-1,\ \bar{g}_{xx}^{(0)}=1,\ \bar{g}^{(0)}_{tx}=0,\ \bar{g}_{tt}^{(2)}=\bar{g}_{xx}^{(2)}=\dfrac{r_+^2+r_-^2}{2},\ \bar{g}_{tx}^{(2)}=r_+r_-.
\ea 

The expectation value of the renormalized stress tensor of $2d$ CFT can be computed by the variation of the action, a Gibbons-Hawking term and the counter-terms in terms of the metric at the $AdS$ boundary as follows~\cite{de Haro:2000xn,Balasubramanian:1999re}:
\ba\label{STR14}
\langle \tilde{T}_{\mu\nu}(x) \rangle =\dfrac{L}{8\pi G_3}(\bar{g}_{\mu\nu}^{(2)}-\bar{g}_{\mu\nu}^{(0)}\bar{g}^{(2)\alpha}{}_{\alpha}).
\ea
Substituting \eqref{MET13} into \eqref{STR14},
\ba
\langle \tilde{T}_{tt} \rangle =\langle \tilde{T}_{xx} \rangle =\dfrac{M L}{2\pi},\ \langle \tilde{T}_{tx}(x) \rangle =\dfrac{J}{2\pi}.
\ea

The stress tensor mentioned above is written in terms of dimensionless parameters. By recovering dimensions $\langle T_{\mu\nu}\rangle =\langle \tilde{T}_{\mu\nu} \rangle /L^2$, we find
\ba\label{ENE22}
&\langle T_{tt}\rangle =\langle T_{xx}\rangle =\dfrac{M}{2\pi L},\ \langle T_{tx} \rangle =\dfrac{J}{2\pi L^2}. 
\ea
Energy is obtained as the integration of energy density \eqref{ENE22} as follows:
\ba\label{MOM25}
&E=\int dx \langle T_{tt} \rangle =M, \\
&P=\int dx \langle T_{t\tilde{x}} \rangle =\dfrac{J}{L},
\ea
where $x$ has a periodicity of $x$. Note that the zero point energy is zero when the mass vanishes. The normalization of energy shows that the anti-de Sitter space with the Poincare coordinate has zero mass.

We must evaluate the grand canonical partition function to obtain the Euclidean path integral
{ \ba
Z=\int [dg]e^{-I_E[g]},
\ea}
where $I_E[g]$ becomes the Euclidean action. Because the classical limit is taken into account, the classical approximation is defined as the steepest descent approximation $Z\sim e^{-I_E[g]}$.

We add Gibbons-Hawking terms and counter-terms to renormalize the action $I_E[g]$ as follows:
{\ba
I_\mathrm{ren}=-\dfrac{1}{16\pi G_3}\int d^3 x\sqrt{g}(R-2\Lambda)+\dfrac{1}{8\pi G_3}\int_{\partial M}d^2x \sqrt{\gamma}\Big(\Theta +\dfrac{1}{L}\Big),
\ea}
where the second term makes the variation principle well-defined.
According to the AdS/CFT correspondence, free energy is temperature times the value of the Euclidean action on the Euclidean continuation of the black hole
{\ba
F&=T I_\mathrm{ren}[{g}],
\ea}, where $1/T$ is the inverse temperature (the Euclidean Killing time). This temperature is obtained as an analytic continuation $J_E\to i J$ and $\tilde{r}_-= i r_-$ of  \eqref{PER13} as follows:
\ba\label{TEM28}
&\beta_0=\dfrac{1}{TL} =\dfrac{2\pi r_+}{r_+^2-r_-^2}.
\ea

Using the metric \eqref{MET2}, free energy becomes
{\ba
\bar{F}=-L^2 M\sqrt{1-\dfrac{J^2}{M^2L^2}}.
\ea}
We can recover dimensions after the coordinate transformation $(t,x,r)\to (t/L,x/L,r/L)$ and $(r_+,r_-)\to (r_+/L,r_-/L)$ (see the measure part). Free energy is represented by energy $M$ and a rotational chemical potential $\Omega$, 
\ba\label{ENE29}
&F&= -M\sqrt{1-\dfrac{J^2}{L^2 M^2}}=M-T s-\Omega J,
\ea
where the spatial volume is $V_x=2\pi L$. According to the Euler relation $M=Ts-PV_x+\mu Q$, free energy is related to pressure $F=-PV_x$. 
The chemical potential is defined as the angular shift of a rotating BTZ black hole at $r=r_+$, namely, {$\Omega =-\frac{4G_3J}{r_+^2}=-\frac{r_-}{r_+L}$. Note that $J$ and $\Omega$ are thermodynamical conjugates. An entropy is given by Bekenstein-Hawking formula $s=\frac{2\pi r_+}{4 G_3}$. The first law of thermodynamics then follows:  $dE=Tds-PdV_x+\mu dQ$. The result of holographic renormalization \eqref{ENE29} is consistent with~\cite{Banados:1992wn,Banados:1992gq}.

Free energy \eqref{ENE29} mentioned above is consistent with the form of the density matrix $e^{-\beta H+i\beta \Omega_E L P}=e^{-\beta H+\beta \Omega L P}$ in the CFT side, where $\Omega_E=i\Omega$ and momentum $P$ is defined in \eqref{MOM25}. That is, the periodicity of a complex coordinate $z=x+i\t_E$ is $2\pi \tau_E= \beta \Omega_EL +i\beta$: $z\sim z+2\pi \tau$.}

The rotating BTZ black hole has the following properties. When the black hole horizon vanishes, temperature \eqref{TEM28} becomes zero. The extreme rotating black hole ($J=Ml$ or $r_+=r_-$) has zero temperature and non-zero entropy. Moreover, the quantum correction to the path integral was computed in~\cite{Carlip:1994gc,Carlip:1992us}. The quantum corrected temperature and entropy were obtained~\cite{Ghosh:1994nz}.

{\section{The double Wick rotated version of a rotating BTZ black hole}}
In this section, we consider the double Wick rotated version of a rotating BTZ black hole. We perform most of the computation in the Euclidean frame and derive the periodicity of this metric. We argue that this spacetime does not have Hawking temperature and entropy in the Lorentzian signature. 

{A rotating asymptotic $AdS_3$ background can be obtained by doing the double Wick rotation(by taking $\tilde{t},\ \tilde{x}$ to $ix,\ it$) on the metric \eqref{MET2}, which is 
\be\label{MET34}
d{s^2} = {L^2}[ - {r^2}{(\frac{{{r_ - }{r_ + }}}{{{r^2}}}dx + dt)^2} + \frac{{({r^2} - r_ - ^2)({r^2} - r_ + ^2)}}{{{r^2}}}d{x^2} + \frac{{{r^2}d{r^2}}}{{({r^2} - r_ - ^2)({r^2} - r_ + ^2)}}].
\ee}
We perform a complete square once again to rewrite \eqref{MET34} in the ADM form as follows:
\ba\label{MET5}
&ds^2=-N^2 dt^2+h_{ab}(dy^a+N^a dt)(dy^b+N^b dt) \nonumber \\
&=l^2 \Big[-\dfrac{(r^2-r_+^2)(r^2-r_-^2)}{(r^2-r_+^2-r_-^2)}dt^2+(r^2-r_+^2-r_-^2)\Big(dx-\dfrac{r_+r_-}{(r^2-r_+^2-r_-^2)}dt\Big)^2 \nonumber \\
&+\dfrac{r^2dr^2}{(r^2-r_+^2)(r^2-r_-^2)}.
\Big]
\ea 
The above formula shows that  a hypersurface $\Sigma_t$ is spanned by $x$ and $r$ at a constant $t$. 
The lapse and the shift are obtained as follows: 
\ba\label{SHI36}
N=\sqrt{\dfrac{(r^2-r_-^2)(r^2-r_+^2)}{r^2-r_-^2-r_+^2}}, \quad N^{\mu}=(0,-\dfrac{r_-r_+}{r^2-r_-^2-r_+^2},0),
\ea
where $N^t=0$ because it is perpendicular to a hypersurface. 

{We take the analytical continuation by letting  $t\to - i\tau_E$ and $r_-\to i\tilde{r}_-$ $(J\to iJ_E)$
\be\label{MET1}
d{s^2} = {L^2}[{r^2}{(\frac{{{{\tilde r}_ - }{r_ + }}}{{{r^2}}}dx - d\tau_E )^2} + \frac{{({r^2} + \tilde r_ - ^2)({r^2} - r_ + ^2)}}{{{r^2}}}d{x^2} + \frac{{{r^2}d{r^2}}}{{({r^2} + \tilde r_ - ^2)({r^2} - r_ + ^2)}}].
\ee}
Correspondingly, we have
\be
\begin{gathered}
  r_ + ^2 = 4{G_3}\left( {M + \sqrt {{M^2} + \frac{{J_E^2}}{{{L^2}}}} } \right), \hfill \\
  r_ - ^2 = 4{G_3}\left( {M - \sqrt {{M^2}{\text{ + }}\frac{{J_E^2}}{{{L^2}}}} } \right). \hfill \\
\end{gathered}
\ee

Note that the metric in \eqref{MET1} becomes equivalent to \eqref{MET8} after interchanging $x$ and $\tau_E$. We can obtain periodicity in spacetime by the double Wick rotation of $x$ and $\tau_E$ directions in \eqref{MET8} and \eqref{PER13}. Alternatively, periodicity can be obtained by requiring the regularity of the metric on a cone spanned by $(x,r)$. The black hole horizon is at $r = {r_ + }$, and the metric near the black hole horizon can be approximated as
\be\label{metric1}
d{s^2} = {L^2}[r_ + ^2{(\frac{{{{\tilde r}_ - }}}{{{r_ + }}}dx - d\tau_E )^2} + \frac{{2(r_ + ^2 + \tilde r_ - ^2)(r - {r_ + })}}{{{r_ + }}}d{x^2} + \frac{{{r_ + }d{r^2}}}{{2(r_ + ^2 + \tilde r_ - ^2)(r - {r_ + })}}].
\ee
By taking
\be
\rho  = L\sqrt {\frac{{2{r_ + }\left( {r - {r_ + }} \right)}}{{r_ + ^2 + \tilde r_ - ^2}}},
\ee
we can have
\be
d{s^2} = {L^2}r_ + ^2{(\frac{{{{\tilde r}_ - }}}{{{r_ + }}}dx - d\tau_E )^2} + {\rho ^{\text{2}}}d{\left( {\frac{{r_ + ^2 + \tilde r_ - ^2}}{{{r_ + }}}x} \right)^2} + d{\rho ^2}.
\ee
To avoid a conical singularity at the black hole horizon, we need to make the following identification
\be
\left( {\tau_E ,x } \right) \sim \left( {\tau_E  + \eta ,x + \zeta } \right)
\ee
with
\be\label{ETA43}
\begin{gathered}
 \eta  =  \frac{{2\pi {{\tilde r}_ - }}}{{r_ + ^2 + \tilde r_ - ^2}}, \hfill \\
  \zeta  = \frac{{2\pi {r_ + }}}{{r_ + ^2 + \tilde r_ - ^2}}. \hfill \\
\end{gathered}
\ee
One can also see periodicity in the ADM decomposition of \eqref{metric1}. When we require the regularity of the metric on a cone spanned by $(\tau_E ,r)$, we obtain periodicity in \eqref{ETA43}. 

We also define the chemical potential, which is equal to the shift \eqref{SHI36} as follows:
\ba
\mu_1=-\dfrac{N^x_E}{L}=\dfrac{r_+}{\tilde{r}_-L}.
\ea
$\eta$ and $\zeta$ satisfy the following equation
\ba
\mu_1L\eta =\zeta . 
\ea 
Compared with the density matrix $\rho =e^{-2\pi \tau_2 H+2\pi \tau_1 i P}$ in the dual CFT, we obtain the complex parameter as follows:
\ba
\tau_1=\dfrac{\zeta}{2\pi}=\dfrac{\mu_1 L\eta}{2\pi},\quad \tau_2 =\dfrac{\eta}{2\pi}. 
\ea 

We then consider an analytic continuation of the double Wick rotated metric \eqref{MET5} to the Lorentz signature. We employ a complex coordinate system $z=x+i \tau_E$. Periodicity of $z$ is given by
\ba
z\sim z+\zeta +i\eta .
\ea
After the analytic continuation $\tilde{r}_-\to -i r_-$ ($\eta \to -i \eta_L$), only the periodicity of spatial directions is obtained as follows: 
\ba\label{PER48}
z\sim z+L_1, \quad \bar{z}\sim \bar{z}+L_2, 
\ea
where the periodicity is defined as
\ba\label{PEB}
L_1\equiv \zeta +\eta_L=\dfrac{2\pi }{\Delta_-}=\dfrac{2\pi}{r_+-r_-},\quad  L_2\equiv \zeta -\eta_L=\dfrac{2\pi }{\Delta_+}=\dfrac{2\pi}{r_++r_-}.
\ea
{Note that the two periodicities of holomorphic and anti-holomorphic sectors are different (see also left and right temperature \eqref{TEMLR} of the rotating BTZ black hole). The periodicity of $\tau_E$ is considered to be infinite in \eqref{PER48}. Thus, the temperature of \eqref{MET34} vanishes. Note that $t=$const and $r=r_+$ surfaces become timelike $g_{xx}=-L^2 r_-^2\le 0$ after the analytic continuation to the Lorentzian frame. It shows that the Bekenstein entropy cannot be defined and the background  \eqref{MET34} has a closed time-like surface~\cite{Banados:1992wn}. Using the stress energy tensor computed in appendix \ref{A} and  changing into the Lorentzian frame $t\to -i\tau_E$ and $r_-\to i\tilde{r}_-$, moreover, we obtain energy density and momentum density as follows:
\ba\label{ENE68}
\epsilon_1=\langle T_{tt} \rangle=-\dfrac{M}{2\pi L},\quad \langle T_{t x} \rangle=-\dfrac{J}{2\pi L^2},\quad \langle T_{xx}\rangle =-\dfrac{M}{2\pi L}.
\ea 
Energy density in \eqref{ENE68} is negative for any $r_-$ (containing $\tilde{r}_-$ in \eqref{RPM}) and analogous to the Casimir energy of dual field theory as described by~\cite{Horowitz:1998ha}. By setting $r_-=0$, actually, the metric becomes the $AdS_3$ soliton. It is dual to field theory at low temperature. These results show that the Lorentzian background \eqref{MET34} is not a black hole.}

\section{Holographic entanglement entropy in the double Wick rotated metric}\label{HEE}
{We derive covariant entanglement entropy in the double Wick rotated version of the rotating BTZ black hole. This background is considered to be dual to CFT with changed periodicity of Euclidean time and spatial direction $(x,t_E)\sim (x+\zeta ,t_E+\eta)$, while the computation is performed in the Lorentzian signature~\eqref{MET34} because the time slice of a time dependent system is well-defined. 

We consider the subsystem $A$ as a strip with a length $\Delta l=x_1-x_2$. The subsystem $B$ is the complement of $A$. The holographic formula is a codimension 2 extremal surface $\gamma_A$ defined as the saddle point of the area function in the Lorentzian spacetime as follows:~\cite{Hubeny:2007xt} 
\ba
S_A=\dfrac{\mbox{Area}(\gamma_A)}{4G_N},
\ea
where $\gamma_A$ has the same boundary as the region $A$ and $\gamma_A$ is homotopic to $A$. The extremal surface corresponds to the space-like geodesic in the bulk connecting points of $\partial A$ when the $3d$ bulk spacetime is considered. {The extremal surface condition is equivalent to the vanishing of null expansions.

The metric \eqref{MET34} is the double Wick rotated version of the rotating BTZ (by taking $\tilde{t},\ \tilde{x}$ to $ix,\ it$). It is locally equivalent to the pure $AdS_3$. The transformation from pure $AdS_3$ to the double Wick rotated metric \eqref{MET34} becomes 
\ba
&w_+=\sqrt{\dfrac{r^2-r_+^2}{r^2-r_-^2}}e^{i(t+x)\Delta_+}=X+T, \nonumber \\
&w_-=\sqrt{\dfrac{r^2-r_+^2}{r^2-r_-^2}}e^{i(t-x)\Delta_-}=X-T, \nonumber \\
&z=\sqrt{\dfrac{r_+^2-r_-^2}{r^2-r_-^2}}e^{i(tr_++xr_-)},
\ea
where $\Delta_{\pm}=r_+\pm r_-$. The pure $AdS$ is written as {$ds^2=\frac{L^2 (dw_+dw_-+dz^2)}{z^2}$}. The spacelike geodesic in pure $AdS_3$ becomes $(X-X_*)^2+z^2=R^2$. The extremal surface is obtained by mapping geodesics in pure $AdS_3$ into the one in the double Wick rotated metric. }

The extremal surface should be on the hypersurface
\ba
\gamma w_+-\gamma^{-1} w_-=const,
\ea
where $\gamma$ means boosts. The subsystem is on a constant time slice of $t_0$. The value of $t$ should be the same. When the extremal surface has two endpoints, $x_1$ and $x_2$, the constraint of the extremal surface becomes 
\ba
\gamma^2 e^{i(t_0+x_1)\Delta_+}-e^{i(t_0-x_1)\Delta_-}=\gamma^2 e^{i(t_0+x_2)\Delta_+}-e^{i (t_0-x_2)\Delta_-}.
\ea

The cut-off and the length of the interval near the $AdS$ boundary become
\ba
&\epsilon =\dfrac{\sqrt{r_+^2-r_-^2}}{r_{\infty}}e^{i(r_+t_0+r_-x_1)}, \nonumber \\
&(\Delta x)^2=\Delta w_+\Delta w_-=(e^{i\Delta_+ (t_0+x_1)}-e^{i\Delta_+(t_0+x_2)})(e^{i\Delta_-(t_0-x_1)}-e^{i\Delta_-(t_0-x_2)}).
\ea

The holographic entanglement entropy becomes the geodesic in $AdS_3$. Substituting the cut-off and the length into the formula of the geodesic, we obtain the entanglement entropy in the double Wick rotated metric as follows:
\ba\label{SA9}
&S_A&=\dfrac{c}{6}\log \Big(\dfrac{(\Delta x)^2}{\epsilon_1\epsilon_2}\Big)=\dfrac{c}{6}\log \Big(\dfrac{4r_{\infty}^2}{\Delta_+\Delta_-}\sin \dfrac{\Delta_+\Delta l}{2}\sin \dfrac{\Delta_-\Delta l}{2}\Big) \nonumber \\
&&=\dfrac{c}{6}\log \Big(\dfrac{(\zeta^2-\eta_L^2)}{\pi^2 \epsilon^2}\sin \dfrac{\pi \Delta l}{\zeta -\eta_L} \sin  \dfrac{\pi \Delta l}{\zeta  +\eta_L}\Big), 
\ea
where $\Delta l =x_1-x_2$ and $r_{\infty}=l^2/\epsilon$. The central charge $c$ is defined as $c=3 L/(2 G_3)$. The entanglement entropy factorizes into the left and right sector 
\ba
S_A=S_R+S_L,
\ea
where
\ba
&S_L=\dfrac{c}{6}\log \Big(\dfrac{L_1}{\pi \epsilon}\sin \dfrac{\pi \Delta l}{L_1} \Big), \nonumber \\
&S_R=\dfrac{c}{6}\log \Big(\dfrac{L_2}{\pi \epsilon}\sin \dfrac{\pi \Delta l}{L_2} \Big).
\ea
Here, $L_1=\zeta +\eta_L$ and $L_2=\zeta -\eta_L$ are the periodicities of the (anti-)holomorphic sector $z$(and $\bar{z}$), respectively (see eq. \eqref{PEB}). The left sector will decouple from the right sector. $S_L$ and $S_R$ are proportional to entanglement entropy of a single interval along a spatial direction with a periodic boundary condition (periodicity $L_1$ and $L_2$). When $r_-=0$ ($\eta_L=0$), $S_L=S_R$ and entanglement entropy $S_A$ becomes \eqref{PER1}.


{\section{The CFT side}}
We periodically identify the two dimensional Euclidean manifold. The partition function of bosons becomes
\ba
Z(\tau ) =\mbox{tr}\Big[\exp (2\pi i \tau_1 P-2\pi \tau_2 H) \Big]=\mbox{tr}(q^{L_0-\frac{c}{24}}\bar{q}^{\tilde{L}_0-\frac{\tilde{c}}{24}}),
\ea
where we have used $q=e^{2\pi i \tau}$. Momentum generates the translation along the $x$ direction  $P=L_0-\tilde{L}_0$ and Hamiltonian generates the translation along the time direction $H=L_0+\tilde{L}_0-\frac{c+\tilde{c}}{24}$. 
To match the gravity dual, the periodicity is chosen as
{\ba
\tau_1=\dfrac{\zeta}{2\pi},\quad \tau_2 =\dfrac{\eta}{2\pi}=-\dfrac{\zeta \Omega_E}{2\pi},
\ea}
where $\Omega_E =1/(\mu_1 L)$. {Because the metric is invariant under the complex conjugation and degenerate for a real $\tau$, we restrict to $\mbox{Im}(\tau)>0$ ($\Omega_E <0$) or vice versa~\cite{Polchinski:1998rq}. }

 This is realized in terms of the following conformal map
\ba\label{CON12}
w=\exp \Big(\frac{i v }{\tau} \Big),
\ea
where the new coordinate $v=x+it_E$ has the periodicity $v\sim v+2\pi \tau$ with modulus $\tau =\frac{\zeta (1-i\Omega_E)}{2\pi}$. Coordinates are transformed as follows:
\ba\label{PER11}
(x,t_E)\sim (x+\zeta , t_E-\zeta \Omega_E).
\ea

One can compare the above result with dual CFT to the rotating BTZ black hole.
Recall that it is obtained by the conformal map $w=\exp (\tilde{w}/\tau)$, where $\tau$ is the same as the above case with $\zeta =\beta$. The new coordinate $\tilde{w}$ has the periodicity $\tilde{w}\sim \tilde{w}+i2\pi \tau$ or 
\ba
(\tilde{x},\tilde{t}_E)\sim (\tilde{x}+\beta \Omega_E,\tilde{t}_E+\beta),
\ea
where $i\tau$ becomes the new modulus. Thus, the periodicity is inversed when it is compared with \eqref{PER11}.

\subsection{Entanglement entropy in the dual CFT}
The value of $\mbox{tr}(\rho_A^n)$ for the reduced density matrix is equivalent to the correlation functions of twisted operators with conformal weights $\delta_n =\bar{\delta}_n=\frac{c}{24}(n-\frac{1}{n})$. 

When CFT lives on $2d$ Euclidean space and the region $A$ is $(u_1\le x\le u_2)$ at a constant time slice, 
\ba
\mbox{tr}(\rho_A^n)=\langle \Phi(u_1) \Phi(u_2)\rangle =c_n\Big(\dfrac{|u_1-u_2|}{\epsilon}\Big)^{-\frac{c}{6}(n-\frac{1}{n})},
\ea 
where the UV cut-off is $\epsilon$ and $c_n$ is a constant.

Under the general conformal transformation, $\mbox{tr}\rho_A^n$ transforms as a 2-point function of primary fields $\Phi$. This implies that one can compute it in different geometries. Under the conformal transformation \eqref{CON12}, the twisted field is transformed as
\ba
\langle \Phi(w_1')\Phi(w_2')\rangle=\Big|\dfrac{\partial w_1^{\prime}}{\partial w_1}\Big|^{-2\delta_n}\Big|\dfrac{\partial w_2^{\prime}}{\partial w_2}\Big|^{-2\delta_n}\langle \Phi(w_1)\Phi(w_2)\rangle,
\ea
where $\partial w'/\partial w =\tau /(iw)$ and $\partial \bar{w}^{\prime}/\partial \bar{w} =-\bar{\tau} /(i\bar{w})$.

We then have 
\ba
\mbox{tr}(\rho_A^n)=\left(\dfrac{\epsilon^2e^{\frac{i}{2}\Big(\frac{w_1'}{\tau}-\frac{\bar{w}_1^{\prime}}{\bar{\tau}}\Big)} e^{\frac{i}{2}\Big(\frac{w_2'}{\tau}-\frac{\bar{w}_2^{\prime}}{\bar{\tau}}\Big)}}{|\tau|^2\Big(e^{i\frac{w_1'}{\tau}}-e^{i\frac{w_2'}{\tau}}\Big)\Big(e^{-i\frac{\bar{w}_1^\prime}{\bar{\tau}}}-e^{-i\frac{\bar{w}_2^\prime}{\bar{\tau}}}\Big)}\right)^{2\delta_n}=\left(\dfrac{\pi^2 \epsilon^2}{\zeta^2 (1+\Omega_E^2)\sin\Big(\frac{\Delta l}{2\tau}\Big)\sin\Big(\frac{\Delta \bar{l}}{2\bar{\tau}}\Big)}\right)^{2\delta_n},
\ea
where the length of the interval becomes $\Delta l=w_1'-w_2'=x_1-x_2$.

By differentiating in terms of $n$, the entanglement entropy is obtained as
\ba\label{SAN68}
&S_A&=-\dfrac{\partial }{\partial n}\log \mbox{tr}(\rho_A^n)\Big|_{n=1}=\dfrac{c}{6}\log \Big(\dfrac{\zeta^2 (1+\Omega_E^2)}{\pi^2 \epsilon^2}\sin \Big(\dfrac{\Delta l}{2\tau}\Big)\sin \Big(\dfrac{\Delta \bar{l}}{2\bar{\tau}}\Big)\Big) \nonumber \\
&&=\dfrac{c}{6}\log \Big(\dfrac{(\zeta^2+\eta^2)}{\pi^2 \epsilon^2}\Big|\sin \dfrac{\pi \Delta l}{\zeta +i\eta}\Big|^2\Big).
\ea
{Note that the entanglement entropy factorizes into two modes, which will decouple in $2d$ CFT. } We perform the analytic continuation $\tilde{r}_-\to - ir_-$ $(\eta \to -i\eta_L)$ and then realize \eqref{SA9} obtained from the gravity dual.

\section{Discussion}
In this paper, the holographic covariant entanglement entropy was analyzed in the double Wick rotated version of a rotating BTZ black hole and agreed with entanglement entropy in the CFT side \eqref{SAN68}. \eqref{SA9} in the Lorentzian signature shows that entanglement entropy factorizes into the left and right sectors. Each sector is proportional to entanglement entropy of a single interval along a compact spatial direction with periodicity $L_1$ and $L_2$ \eqref{PEB}. The constraint of the causality annoys in the Lorentzian signature as mentioned below. However, analysis on the CFT side implies that it will also work for any places where the extremal surface probes the interior of the background \eqref{MET34}.

The entanglement entropy has also been computed on the CFT side. The entanglement entropy was obtained by a conformal transformation \eqref{CON12} of correlation functions of twisted operators in the analogue of~\cite{Hubeny:2007xt}. The entanglement entropy factorized into two modes, which would decouple like the one in the gravity dual. {Because a cylinder can be considered as the infinite limit of a torus periodicity, our result should be realized from entanglement entropy on the torus~\cite{Herzog:2013py,Kim:2017ghc}. The modular transformation of entanglement entropy on the torus will also be interesting. When the chemical potential conjugate to momentum is zero, $\tau\to -1/\tau$ transforms from zero temperature to high temperature. The modular transformation analysis with $\tau_1\neq 0$ should be applicable for ours.}  

The double Wick rotated geometry may have problems with causality in the Lorentzian frame. We make use of a Killing vector to analyze the causal structure of the background. A Killing vector $\partial_x$ has the norm $\xi\cdot \xi = r^2-r_-^2-r_+^2$, which is spacelike for $r^2\ge r_+^2+r_-^2$. However, it becomes timelike for  $r^2\le r_+^2+r_-^2$.~\footnote{ In addition,  $t=$const and $r=r_+$ surfaces become timelike $g_{xx}=-L^2 r_-^2\le 0$. }  {It shows that the double Wick rotated geometry has a closed time-like curve~\cite{Banados:1992wn}. It will be possible to explore how the law of physics allows a closed time-like curve~\cite{Friedman:1990xc}.} Second, the background is not a black hole in the Lorentzian signature because it does not have Hawking temperature, which vanishes after the analytic continuation $\tilde{r}_-\to -i r_-$ (see \eqref{ETA43}).  It also has negative energy, unlike thermal field theory. This negative energy will correspond to Casimir energy in the CFT side~\cite{Horowitz:1998ha}.

In the Euclidean signature, on the other hand, the periodicity of Euclidean time and spatial direction for the double Wick rotated version of the rotating BTZ black hole is changed as in \eqref{ETA43}. This background has properties of rotating $AdS$ black holes, such as Hawking temperature by requiring regularity at the black hole horizon. The black hole entropy will appear in this frame. It will be interesting to analyze the thermodynamics of the double Wick rotated geometry. The entanglement entropy~\eqref{SAN68} in the Euclidean signature will realize the results of dual CFT to black holes. 

{It will be interesting when we identify the periodicity of both a rotating BTZ and the double Wick rotated metric in the Euclidean frame  as follows: 
\ba
(x,\tau_E)\sim (x+\zeta, \tau_E+\eta ).
\ea
 Substituting $\eta=\beta_0$ and $\zeta=\phi_0$ in~\eqref{PER13} and using \eqref{ETA43}, $r_+$ is interchanged with $\tilde{r}_-$ between two backgrounds. The observable of both theories will then be consistent because both theories are  CFT with the same periodicity. We call the holographic stress energy tensor of a rotating BTZ (the double Wick rotated metric) $\langle T_{\mu\nu} \rangle^{BTZ}$ \eqref{ENE22} ($\langle T_{\mu\nu} \rangle^{DWR}$ \eqref{ENE68}), respectively. We find that both theories have the same form of the holographic stress tensor as follows:
\ba
&\langle T_{tt} \rangle^{BTZ} =\langle T_{tt} \rangle^{DWR}=\dfrac{\pi L (\eta^2-\zeta^2)}{4G_3  (\zeta^2+\eta^2)^2}. \nonumber \\
&\langle T_{tx} \rangle^{BTZ} =\langle T_{tx} \rangle^{DWR}=-\dfrac{i \pi L \zeta \eta}{2 G_3 (\zeta^2+\eta^2)^2},
\ea 
Physical quantities such as entropy density and free energy can also be computed. Free energy should be consistent with the form of the density matrix on the CFT side. It will be interesting to proceed in this direction furthermore.}

\section*{Acknowledgments}
We would like to thank B. S. Kim for collaboration in the initial stage of this project and for helpful discussions. 
MF would like to thank S. He, S. Lin, and J. Sun for discussions and comments. We also thank T. Takayanagi for valuable discussion and comments.

\appendix
\section{Holographic stress energy tensor}\label{A}
In this appendix, we compute the holographic stress tensor for the double Wick rotated metric. 
We compute the holographic stress energy tensor in the FG coordinate $ds^2=\frac{L^2}{\rho^2}(d\rho^2+\bar{g}_{ij}^Edx^idx^j)$, where $\bar{g}^E_{ij}=\bar{g}^{E(0)}_{ij}+\bar{g}^{E(2)}_{ij}\rho^2+\dots$. Metric for \eqref{MET5} becomes
\ba
&\bar{g}_{\tau_E\tau_E}^{E(0)}=1,\quad \bar{g}_{xx}^{E(0)}=1, \quad \bar{g}_{\tau_Ex}^{E(0)}=0, \nonumber \\ 
&\bar{g}_{\tau_E\tau_E}^{E(2)}=\dfrac{r_+^2-\tilde{r}_-^2}{2},\quad \bar{g}_{xx}^{E(2)}=\dfrac{-r_+^2+\tilde{r}_-^2}{2},\quad \bar{g}^{E(2)}_{\tau_E x}=-r_+\tilde{r}_-. \label{BOU22}
\ea 
The boundary stress tensor becomes
\ba\label{STR23}
\langle \tilde{T}_{\mu\nu}^E(x)\rangle =\dfrac{L}{8\pi G_3}(\bar{g}_{\mu\nu}^{(2)}-\bar{g}_{\mu\nu}^{(0)}\bar{g}^{2\alpha}{}_{\alpha}). 
\ea
Substituting \eqref{BOU22} into \eqref{STR23} and performing the coordinate transformation into $(\tau_E', x')=(L\tau_E ,L x )$, we obtain the stress energy tensor in the dual CFT as follows:
\ba
\langle T^E_{\tau_E\tau_E} \rangle=\dfrac{M_E}{2\pi L},\quad \langle T_{\tau_E x} \rangle=-\dfrac{J_E}{2\pi L^2},\quad \langle T_{xx}\rangle =-\dfrac{M_E}{2\pi L}.
\ea


\end{document}